\documentclass[12pt]{amsart}
\usepackage{amssymb}
\usepackage{color}
\usepackage{amsmath,epic,curves}
\usepackage[english]{babel}
\usepackage{graphicx}
\usepackage{comment}
\usepackage{appendix}
\begin{document}
\baselineskip 6.0 truemm
\parindent 1.5 true pc

\newcommand\lan{\langle}
\newcommand\ran{\rangle}
\newcommand\tr{{\text{\rm Tr}}\,}
\newcommand\ot{\otimes}
\newcommand\ol{\overline}
\newcommand\join{\vee}
\newcommand\meet{\wedge}
\renewcommand\ker{{\text{\rm Ker}}\,}
\newcommand\image{{\text{\rm Im}}\,}
\newcommand\id{{\text{\rm id}}}
\newcommand\tp{{\text{\rm tp}}}
\newcommand\pr{\prime}
\newcommand\e{\epsilon}
\newcommand\la{\lambda}
\newcommand\inte{{\text{\rm int}}\,}
\newcommand\ttt{{\text{\rm t}}}
\newcommand\spa{{\text{\rm span}}\,}
\newcommand\conv{{\text{\rm conv}}\,}
\newcommand\rank{\ {\text{\rm rank of}}\ }
\newcommand\re{{\text{\rm Re}}\,}
\newcommand\ppt{\mathbb T}
\newcommand\rk{{\text{\rm rank}}\,}

\newcommand{\bra}[1]{\langle{#1}|}
\newcommand{\ket}[1]{|{#1}\rangle}

\newcommand\bfi{{\bf i}}
\newcommand\bfj{{\bf j}}
\newcommand\bfk{{\bf k}}
\newcommand\bfl{{\bf l}}
\newcommand\bfzero{{\bf 0}}
\newcommand\bfone{{\bf 1}}

\title{Construction of three qubit genuine entanglement with
bi-partite positive partial transposes}

\author{Kil-Chan Ha and Seung-Hyeok Kye}
\address{Faculty of Mathematics and Statistics, Sejong University, Seoul 143-747, Korea}
\address{Department of Mathematics and Institute of Mathematics, Seoul National University, Seoul 151-742, Korea}

\thanks{KCH and SHK were partially supported by NRFK grant 2014-010499 and  NRFK grant 2015R1D1A1A02061612 respectively.}



\begin{abstract}
We construct tri-qubit genuinely entangled states which have
positive partial transposes with respect to bi-partition of systems.
These examples disprove a conjecture \cite{novo}
which claims that PPT
mixtures are necessary and sufficient for biseparability of three
qubits.
\end{abstract}

\maketitle

\section{Introduction}

The notion of entanglement is now considered as one of the key
resources in quantum information and quantum computation theory, and
it is an important research topic to distinguish entanglement from
separability. It is known \cite{{choi-ppt},{peres}} that a separable
state is of PPT, that is, the partial transpose of a separable state
is still positive. This PPT condition  is also sufficient for
separability in the low dimensional bi-partite cases, like bi-qubit
and qubit-qutrit systems \cite{horo-1,woronowicz}. In the tri-qubit
case, it was conjectured in \cite{novo} that bi-separability is
equivalent to being a PPT mixture. This conjecture is supported by
several classes of states; graph-diagonal states \cite{guhne11},
permutationally invariant states \cite{novo} and {\sf X}-shaped
states \cite{han_kye_X_state}. The purpose of the paper is to
construct analytic examples to disprove this conjecture for general cases.
Note that numerical evidence for such states has been mentioned
in the supplemental material of \cite{toth}.

A multi-partite state $\varrho$ in $\bigotimes_{i=1}^n M_{d_i}$ is {\sl (genuinely) separable} if it is the convex combination
of pure product states onto product vectors, where $M_d$ denotes the algebra of all $d\times d$ matrices over the complex field.
 For a given bipartition $S\sqcup T$ of the set $[n]:=\{1,2,\dots,n\}$ of indices,
$\varrho$ is said to be {\sl $S$-$T$ bi-separable} if it is separable
as a bi-partite state in $(\bigotimes_{i\in S}M_{d_i})\ot(\bigotimes_{i\in T}M_{d_i})$,
and just {\sl bi-separable} if it is the convex combination of $S$-$T$ bi-separable states through bi-partitions
$S\sqcup T$. A multi-partite state is called
{\sl genuinely entangled} if it is not bi-separable.

The notion of PPT mixture is defined in the exactly same way. For a given bi-partition $S\sqcup T$,
a multi-partite state $\varrho$ in $\bigotimes_{i=1}^n M_{d_i}$ is said to be a {\sl $S$-$T$ bi-PPT state} if
it is of PPT as a bi-partite state in $(\bigotimes_{i\in S}M_{d_i})\ot(\bigotimes_{i\in T}M_{d_i})$,
and  is called a {\sl PPT mixture} \cite{jungn} if it is the convex combination of
$S$-$T$ bi-PPT states through bi-partitions $S\sqcup T$.

We begin our construction with the example of a nonextendible
positive linear map from $M_2$ into $M_4$ given by Woronowicz
\cite{woronowicz-1}. In the next section, we show that the Choi
matrix $W$ of this map is a tri-qubit genuine entanglement witness:
$W$ is a non-positive self-adjoint $8\times 8$ matrix satisfying
\begin{equation}\label{witness}
\lan \varrho,W\ran:=\tr(W\varrho^\ttt)\ge 0
\end{equation}
for every tri-qubit bi-separable state $\varrho$. Nonextendibility
of the map implies the exposedness
\cite{{woro_letter},{chrus_exposed}}, which implies again that both
$W$ and its partial transpose $W^\Gamma$ have the spanning property,
where we take partial transpose with respect to the bi-partition
$A$-$BC$ for $M_A\ot M_B\ot M_C$. See Theorem 8.5 in
\cite{kye_ritsu}. Therefore, there exist $A$-$BC$ bi-separable
states $\varrho_1$ with the properties:
\begin{itemize}
\item
$\lan \varrho_1,W\ran =0$,
\item
both $\varrho_1$ and $\varrho_1^\Gamma$ have the full rank.
\end{itemize}
We consider the convex set $\mathbb S$ consisting of all $A$-$BC$ bi-separable states, and
the face of $\mathbb S$ consisting of all $\varrho\in\mathbb S$ with
$\lan\varrho,W\ran=0$. This face turns out to be maximal, and every interior point of this face actually satisfies these conditions.

Now, we take $\varrho_0:=\frac 18 {\text{\rm Id}}_8$ which is located at the center of the whole things, and consider the line segment
$$
\varrho_\lambda=(1-\lambda)\varrho_0+\lambda\varrho_1
$$
from $\varrho_0$ to $\varrho_1$. We may extend this line segment to get $A$-$BC$ bi-PPT states $\varrho_\lambda$
with $\lambda >1$. Then, this is obviously a tri-qubit PPT mixture which violates the inequality (\ref{witness}).
This is the outline of our construction.
In the Section 3, we construct an analytic example which is based on the example of a separable state $\varrho_1$
with the above conditions constructed in \cite{ha_kye_woro_exam}, where Woronowicz map has been generalized with
several parameters.

The authors are grateful to Otfried G\" uhne for bringing their attention to the reference \cite{toth}.

\section{Genuine entanglement witnesses}

In order to express entries of multi-partite entanglement witnesses in the tensor product $\bigotimes_{i=1}^n M_{d_i}$,
we will use multi-indices. A multi-index $\bfi $ on a nonempty subset $S$  of $[n]=\{1,2,\dots,n\}$
is formally a function from $S$ into nonnegative integers with $0\le {\bfi}(i)<d_i$ ($i\in S$), which will be
denoted by a string of integers in the obvious sense.
For a multi-index ${\bfi}=i_1i_2\dots i_{\# S}$ on $S$, we also use the notation
$$
|{\bfi}\ran=|i_1\ran\ot |i_2\ran\ot \cdots\ot |i_{\# S}\ran,
$$
where $\#S$ denotes the cardinality of $S$.

For a given bi-partition $S\sqcup T=[n]$, any entanglement witness $W$ in $\bigotimes_{i\in [n]} M_{d_i}$
is written by
$$
W=\sum_{{\bfi},{\bfj}\in I_S} |{\bfi}\ran\lan {\bfj}| \ot W[{\bfi},{\bfj}]\in
\left(\bigotimes_{i\in S} M_{d_i}\right)\otimes \left(\bigotimes_{i\in T} M_{d_i}\right)
$$
in a unique way,
where $I_S$ denotes the set of all multi-indices on the set $S$.
For given multi-indices ${\bfk},{\bfl}$ on $T$,
the $({\bfk},{\bfl})$-entry $W[{\bfi},{\bfj}]_{{\bfk},{\bfl}}$ of $W[{\bfi},{\bfj}]\in \bigotimes_{i\in T} M_{d_i}$ is given by
$$
W[{\bfi},{\bfj}]_{{\bfk},{\bfl}}=W_{{\bfi} \diamond {\bfk}, {\bfj} \diamond {\bfl}},
$$
where ${\bfi}\diamond {\bfk}$ is the multi-index on $[n]$ given by
$({\bfi} \diamond {\bfk})(i )={\bfi}(i )$ for $i \in S$ and
$({\bfi} \diamond {\bfk})(i )={\bfk}(i )$ for $i \in T$.
Since the set $\{|{\bfi}\ran\lan {\bfj}| : {\bfi}, {\bfj} \in I_S \}$ plays the role of matrix units
for $\bigotimes_{i\in S} M_{d_i}$, we can define the linear map
\begin{equation}\label{st_map}
\phi_W^{S,T}: |{\bfi}\ran\lan {\bfj}| \in \bigotimes_{i\in S} M_{d_i}\to W[{\bfi},{\bfj}]
\in \bigotimes_{i\in T} M_{d_i}, \qquad {\bfi}, {\bfj} \in I_S.
\end{equation}
If $n=2$ and $S=\{1\}$, $W$ is the usual Choi matrix \cite{choi75-10} of the linear map $\phi_W^{\{1\},\{2\}}$ from $M_{d_1}$
into $M_{d_2}$.
It was shown in \cite{han_kye_X_state} that $W$ satisfies the inequality (\ref{witness}) for every bi-separable state $\varrho$
if and only if the map
$\phi_W^{S,T}$ is positive for every nonempty subset $S$ with $\#S\le \frac n2$.

Now, the Choi matrix of the Woronowicz map \cite{woronowicz-1} is given by
$$
W=\left(
\begin{matrix}
4 & -2  &\cdot&\cdot&-2 & \cdot  &\cdot&\cdot\\
-2& 2 &\cdot &\cdot&2& \cdot &\cdot &\cdot\\
\cdot&\cdot &\cdot&\cdot&\cdot&1 &\cdot&\cdot\\
\cdot&\cdot&\cdot&4&\cdot&\cdot&-2&\cdot\\
-2 & 2  &\cdot&\cdot&3 & \cdot  &\cdot&\cdot\\
\cdot& \cdot &1 &\cdot&\cdot& \cdot &\cdot &\cdot\\
\cdot&\cdot &\cdot&-2&\cdot&\cdot &2&-1\\
\cdot&\cdot&\cdot&\cdot&\cdot&\cdot&-1&2
\end{matrix}
\right),
$$
where $\cdot$ denotes $0$, if we endow multi-indices with the lexicographic order.
The linear map $\phi_W^{A,BC}$ is the original positive map of Woronowicz \cite{woronowicz-1} which
sends
$X=\left(\begin{matrix}x&y\\z&w\end{matrix}\right)$ to
$$
\phi_W^{A,BC}(X)=
\left(
\begin{matrix}
4x-2(y+z)+3w & -2x+2z  &\cdot&\cdot\\
-2x+2y& 2x &z &\cdot\\
\cdot&y &2w&-2z-w\\
\cdot&\cdot&-2y-w&4x+2w
\end{matrix}
\right).
$$
On the other hand, the linear maps $\phi_W^{B,CA}$ and $\phi_W^{C,AB}$ send $X=\left(\begin{matrix}x&y\\z&w\end{matrix}\right)$ to
$$
\phi_W^{B,CA}(X)=
\left(
\begin{matrix}
4x & -2x  &-2x&z\\
-2x& 2x+4w &2x-2w &\cdot\\
-2x&2x-2w &3x+2w&-w\\
y&\cdot&-w&2w
\end{matrix}
\right)
$$
and
$$
\phi_W^{C,AB}(X)=
\left(
\begin{matrix}
4x-2(y+z)+2w & \cdot  &-2x+2z&\cdot\\
\cdot& 4w &y &-2z\\
-2x+2y&z &3x&\cdot\\
\cdot&-2y&\cdot&2x-y-z+2w
\end{matrix}
\right),
$$
respectively.

Now, we proceed to show that $\phi_W^{B,CA}$ and $\phi_W^{C,AB}$ are positive maps from $M_2$ into $M_4$.
To do this, it suffices to show that the images of
$P_\alpha=\left(\begin{matrix}1&\bar\alpha\\ \alpha&|\alpha|^2\end{matrix}\right)$
are positive for each complex number $\alpha$. For the case of $\alpha=0$,
it is easy to show that both $\phi_W^{B,CA}(P_0)$
and $\phi_W^{C,AB}(P_0)$ are positive semi-definite.

For the case of $\alpha\neq 0$, we consider the determinant $\Delta^B_k(\alpha)$ ($\Delta^C_k(\alpha)$, respectively)
of left-upper $k\times k$ submatrix of $\phi_W^{B,CA}(P_{\alpha})$ ($\phi_W^{C,AB}(P_{\alpha})$, respectively)
for each $k=1, 2, 3, 4$. By direct computation, we have
\[
\begin{aligned}
\Delta^B_1(\alpha) =&\,4>0,\\
\Delta^B_2(\alpha) =&\,4(1+4|\alpha|^2)>0,\\
\Delta^B_3(\alpha) =&\,4(1+14|\alpha|^2+4|\alpha|^4)>0,\\
\Delta^B_4(\alpha) =&\,6|\alpha|^2\bigl(1+14|\alpha|^2+2|\alpha|^4+2|\alpha|^2(\alpha+\bar \alpha)\bigr)\\
\ge &\,6|\alpha|^2(1+14|\alpha|^2-4|\alpha|^3+2|\alpha|^4)=
6|\alpha|^2\bigl(1+|\alpha|^2(14-4|\alpha|+2|\alpha|^2)\bigr)> 0,\\
\Delta^C_1(\alpha) =&\,2|\alpha-1|^2+2>0,\\
\Delta^C_2(\alpha) =&\,8|\alpha|^2(|\alpha-1|^2+1)> 0,\\
\Delta^C_3(\alpha) =&\,2|\alpha|^2(3|\alpha-1|^2+11)> 0,\\
\Delta^C_4(\alpha) =&\,2|\alpha|^2\bigl(12-16(\alpha+\bar \alpha)
  +3(\alpha^2+\bar \alpha^2)-9|\alpha|^2(\alpha+\bar \alpha)+36|\alpha|^2+6|\alpha|^4\bigr)\\
=&\,2|\alpha|^2\left(\left|3\alpha+\bar \alpha-\frac 9 4 |\alpha|^2\right|^2+\frac {15}{16}|\alpha|^4
  +26|\alpha|^2-16(\alpha+\bar \alpha)+12\right)\\
\ge&\,2|\alpha|^2\left(\left|3\alpha+\bar \alpha-\frac 9 4 |\alpha|^2\right|^2
  +\frac {15}{16}|\alpha|^4\right)+4|\alpha|^2\bigl(13|\alpha|^2-16|\alpha|+6\bigr)>0.
\end{aligned}
\]
The above results imply that $\phi_W^{B,CA}(P_\alpha)$ and $\phi_W^{C,AB}(P_\alpha)$ are positive definite for each $\alpha\neq 0$.
Consequently, we conclude that $\phi_W^{B,CA}$ and $\phi_W^{C,AB}$ are positive maps from $M_2$ into $M_4$,
and so $W$ is a tri-qubit genuine entanglement witness.

\section{Construction}
Since $W$ and its partial transpose $W^{\Gamma}$ with respect to the bi-partition $A$-$BC$ have the spanning property,
we can choose finitely many product vectors $|z_i\rangle=|x_i\otimes y_i\rangle:=|x_i\rangle \otimes |y_i\rangle$
with the properties:
\begin{itemize}
\item
$\langle x_i\otimes y_i|W|x_i\otimes y_i\rangle = \langle \bar x_i\otimes y_i|W^{\Gamma}|\bar x_i\otimes y_i\rangle=0$,
\item
both $\{|x_i\otimes y_i\rangle\}$ and $\{|\bar x_i\otimes y_i\rangle\}$ span $\mathbb C^2\ot\mathbb C^4$.
\end{itemize}
For example, we take the following eight complex numbers
\[
\begin{aligned}
&\alpha_1=1,\quad \quad \quad \quad \, \alpha_2=i,\quad \quad \quad \quad \  \alpha_3=-1,\quad \quad \quad \quad \,  \alpha_4=-i,\\
&\alpha_5=\sqrt{2}\,(1+i),\ \alpha_6=\sqrt{2}\,(1-i),\ \alpha_7=-\sqrt{2}\,(1-i),\ \alpha_8=-\sqrt{2}\,(1+i),
\end{aligned}
\]
and define  $|x_k\rangle \in \mathbb C^2$ and $|y_k\rangle \in \mathbb C^4$ by
\[
\begin{aligned}
&|x_k\rangle =(1,\bar \alpha_k)^{\rm t}, \\
&|y_k\rangle =\bigl(2\alpha_k-2\alpha_k^2,\,
4\alpha_k -2\alpha_k^2 -2|\alpha_k|^2+3\alpha_k |\alpha_k|^2,\,-4-2|\alpha_k|^2,\,-2\bar \alpha_k-|\alpha_k|^2\bigr)^{\rm t},
\end{aligned}
\]
for each $k=1,\,2,\,\cdots,\,8$.
See the Ref. \cite{ha_kye_woro_exam} for systematic construction of such product vectors in more general cases.


Now, we consider the following two states
\[
\sigma_1=\dfrac{1}{848}\sum_{k=1}^4 |z_k\rangle \langle z_k|,\quad
\sigma_2=\dfrac{1}{28160}\sum_{k=5}^8 |z_k\rangle \langle z_k|,
\]
and define the state $\varrho_1$ by
\[
\varrho_1=\dfrac{1}{12}\sigma_1+\dfrac{11}{12}\sigma_2=\begin{pmatrix}
\frac{23}{1696} & \frac{17}{530} & 0 & 0 & -\frac{21}{2120} & -\frac{21}{2120} & 0 & -\frac{43}{6360}\\
\\
\frac{17}{530} & \frac{731}{4240} &  \frac{73}{4240} & \frac{21}{4240} &  -\frac{87}{1060} & -\frac{153}{1060} &  0 &-\frac{43}{6360}\\
\\
0 & \frac{73}{4240} & \frac{279}{8480} & \frac{73}{8480} & -\frac{73}{4240} & -\frac{247}{2120} &  0 &  0\\
\\
0 & \frac{21}{4240} & \frac{73}{8480} & \frac{13}{2120} & -\frac{21}{4240} & -\frac{111}{4240} & \frac{73}{4240} & \frac{21}{4240}\\
\\
-\frac{21}{2120} & -\frac{87}{1060} & -\frac{73}{4240} & -\frac{21}{4240} & \frac{19}{424} & \frac{227}{2120} & 0 & 0\\
\\
-\frac{21}{2120} & -\frac{153}{1060} & -\frac{247}{2120} & -\frac{111}{4240} & \frac{227}{2120} & \frac{2639}{4240} & \frac{29}{530} & \frac{37}{2120}\\
\\
0 & 0 &  0 & \frac{73}{4240} & 0 & \frac{29}{530} & \frac{189}{2120} & \frac{29}{1060}\\
\\
-\frac{43}{6360} & -\frac{43}{6360} & 0 & \frac{21}{4240} & 0 & \frac{37}{2120} & \frac{29}{1060} & \frac{79}{4240}
\end{pmatrix}.
\]
Then $\langle \varrho_1,W\rangle=0$ by construction.

On the other hand, the characteristic polynomial $f_{\varrho_1}(x)$ of $\varrho_1$ is given by
\[
\begin{aligned}
f_{\varrho_1}(x)=&\, 882 - 11102415 x + 33652192213 x^2 - 14743193830128 x^3 \\ &+
 2042243130647296 x^4 - 99440801962659840 x^5
+ 1345976609197260800 x^6 \\& \quad - 5957113683640320000 x^7 +
 5957113683640320000 x^8,
\end{aligned}
\]
and we can show that all of the roots of $f_{\varrho_1}(x)=0$ are positive real numbers with the smallest root
$\lambda_{\varrho_1}\approx 1.23\times 10^{-4}$.
We can also show that
$\lambda_{\varrho_1^{\Gamma}}\approx 3.5\times 10^{-4}$ is the smallest root of the characteristic equation
$f_{\varrho_1^{\Gamma}}(x)=0$ of $\varrho_1^{\Gamma}$, where
\[
\begin{aligned}
f_{\varrho_1^{\Gamma}}(x)=&\, 2025 - 10945449 x + 18736999508 x^2 - 12463670543488 x^3 \\
&+  3063341671496192 x^4 - 162870055626219520 x^5 +
 2691953218394521600 x^6 \\
&\quad - 11914227367280640000 x^7 +
 11914227367280640000 x^8.
\end{aligned}
\]
Consequently, $\varrho_1$ is an $A$-$BC$ bi-PPT state with $\textrm{rank}(\varrho_1)=\textrm{rank}(\varrho_1^{\Gamma})$=8.

Finally, we consider
$$
\varrho:=\varrho_1-\lambda_{\varrho_1}\textrm{Id}_8.
$$
Since $0<\lambda_{\varrho_1}<\lambda_{\varrho_1^{\Gamma}}$,
we see that
$\varrho$ is an unnormalized  $A$-$BC$ bi-PPT state with $\textrm{rank}(\varrho)=7$ and $\textrm{rank}(\varrho^{\Gamma})=8$.
Furthermore, we see that $\langle \varrho, W\rangle=-17\lambda_{\varrho_1}<0$ with the tri-qubit genuine entanglement witness $W$
in the previous section.
Therefore, we conclude that $\frac{1}{\textrm{Tr}(\varrho)} \varrho$ is an example of a tri-qubit PPT mixture which is not bi-separable.

In conclusion, we have constructed a genuinely entangled state which is a PPT mixture.


\begin{thebibliography}{99}

\bibitem{choi75-10}
M.-D. Choi, \it Completely positive linear maps on complex matrices,
\rm Linear Alg. Appl. \bf 10 \rm (1975), 285--290.

\bibitem{choi-ppt}
M.-D. Choi, \it Positive linear maps, \rm Operator Algebras and
Applications (Kingston, 1980), pp. 583--590, Proc. Sympos. Pure
Math. Vol 38. Part 2, Amer. Math. Soc., 1982.

\bibitem{chrus_exposed}
D. Chru\'{s}ci\'{n}ski and G. Sarbicki, \it Exposed positive maps: a
sufficient condition, \rm J. Phys. A: Math. Theor. \bf 45 \rm
(2012), 115304.

\bibitem{guhne11}
O. G\" uhne, B. Jungnitsch, T. Moroder and Y. S. Weinstein,
\it Multiparticle entanglement in graph-diagonal states: Necessary and sufficient
conditions for four qubits,
\rm Phys. Rev. A {\bf 84} (2011), 052319.

\bibitem{ha_kye_woro_exam}
K.-C. Ha and S.-H. Kye,
\it Construction of exposed indecomposable positive linear maps between matrix algebras,
\rm preprint. arXiv:1410.5545

\bibitem{han_kye_tri}
K. H. Han and S.-H, Kye,
\it Various notions of positivity for bi-linear maps and applications to tri-partite entanglement,
\rm J. Math. Phys. {\bf 57} (2016), 015205.

\bibitem{han_kye_X_state}
K. H. Han and S.-H, Kye,
\it Construction of multi-qubit optimal genuine entanglement witnesses,
\rm preprint. arXiv:1510.03620

\bibitem{horo-1}
M. Horodecki, P. Horodecki and R. Horodecki,
\it Separability of mixed states: necessary and sufficient conditions,
\rm Phys. Lett. A \bf 223 \rm (1996), 1--8.

\bibitem{jungn}
B. Jungnitsch, T. and O. G\" uhne,
\it Taming Multiparticle Entanglement,
\rm Phys. Rev. Lett. {\bf 106} (2011), 190502.



\bibitem{kye_ritsu}
S.-H. Kye,
\it Facial structures for various notions of positivity and applications to the theory of entanglement,
\rm Rev. Math. Phys. \bf 25 \rm (2013), 1330002.

\bibitem{kye_multi_dual}
S.-H. Kye,
\it Three-qubit entanglement witnesses with the full spanning properties,
\rm J. Phys. A: Math. Theor. {\bf 48} (2015), 235303.

\bibitem{novo}
L. Novo, T. Moroder and O. G\" uhne,
\it Genuine multiparticle entanglement of permutationally invariant states,
\rm Phys. Rev A {\bf 88} (2013), 012305.

\bibitem{peres}
A. Peres,
\it Separability Criterion for Density Matrices,
\rm Phys. Rev. Lett. \bf 77 \rm (1996), 1413--1415.

\bibitem{toth}
G. T\' oth, T. Moroder and O. G\" uhne,
\it  Evaluating convex roof entanglement measures,
\rm Phys. Rev. Lett. {\bf 114} (2015), 160501:
Supplemental Material, arXiv 1409.3806


\bibitem{woronowicz}
S. L. Woronowicz,
\it Positive maps of low dimensional matrix algebras,
\rm Rep. Math. Phys. \bf 10 \rm (1976), 165--183.

\bibitem{woronowicz-1}
S. L. Woronowicz,
\it  Nonextendible positive maps,
\rm Commun. Math. Phys. \bf 51 \rm (1976), 243--282.

\bibitem{woro_letter}
S. L. Woronowicz, \it Exposed positive maps, \rm private
communication, July 2011.

\end{thebibliography}
\end{document}